\RequirePackage{lineno}
\documentclass[aps,prl,twocolumn,showpacs,superscriptaddress,groupedaddress]{revtex4}  % for review and submission
\usepackage{graphicx}  % needed for figures
\usepackage{dcolumn}   % needed for some tables
\usepackage{bm}        % for math
\usepackage{amssymb}   % for math
\usepackage{amsmath}   % for math
\usepackage{xspace}

\newcommand{\dzero}    {D0\xspace}
\newcommand{\ttbar}    {\ensuremath{t\bar{t}}\xspace}
\newcommand{\ppbar}    {\ensuremath{p\bar{p}}\xspace}
\newcommand{\qqbar}    {\ensuremath{q\bar{q}}\xspace}
\newcommand{\ee}       {\ensuremath{ee}\xspace}
\newcommand{\emu}      {\ensuremath{e\mu}\xspace}
\newcommand{\mumu}     {\ensuremath{\mu\mu}\xspace}
\newcommand{\zjj}      {\ensuremath{Z+\rm 2~jets}\xspace}
\newcommand{\ztautauemujj}
           {\ensuremath{(Z\rightarrow\tau^+\tau^-\rightarrow e^+\nu_e \mu^- \bar{\nu}_{\mu})+\rm 2~jets}\xspace}

\newcommand{\zee}{\ensuremath{Z\rightarrow e^+e^-}\xspace}
\newcommand{\zmumu}{\ensuremath{Z\rightarrow\mu^+\mu^-}\xspace}
\newcommand{\ztautau}{\ensuremath{Z \rightarrow \tau^+ \tau^-}\xspace}
\newcommand{\ztautauemu}{\ensuremath{Z \rightarrow \tau^+ \tau^-\rightarrow e^+ \nu_e  \mu^-\bar{\nu}_{\mu}}\xspace}
\newcommand{\zll}{\ensuremath{Z/\gamma^{\ast}\to\ell^+\ell^-}\xspace}
\newcommand{\wtaunu}{\ensuremath{W^+\rightarrow \tau^+ \nu_{\tau}}\xspace}
\newcommand{\taulnunu}{\ensuremath{\tau^+ \rightarrow \ell^+ \nu_{\ell} \bar{\nu}_{\tau}}\xspace}
\newcommand{\wlnu}{\ensuremath{W^+\rightarrow \ell^+ \nu_{\ell}}\xspace}

\newcommand{\pt}       {\ensuremath{p_{T}}\xspace}
\newcommand{\ptttbar}  {\ensuremath{p_{{T}}^{\ttbar}}\xspace}
\newcommand{\mt}       {\ensuremath{m_{t}}\xspace}
\newcommand{\pevt}     {\ensuremath{P_{\rm evt}}\xspace}
\newcommand{\pttbar}   {\ensuremath{P_{\ttbar}}\xspace}
\newcommand{\ptot}     {\ensuremath{L_{\rm tot}}\xspace}
\newcommand{\sigmaobs} {\ensuremath{\sigma_{\rm obs}}\xspace}

\newcommand{\pzjj}     {\ensuremath{P_{\zjj}}\xspace}
\newcommand{\fttbar}   {\ensuremath{f_{\ttbar}}\xspace}
\newcommand{\fulldecay}{\ensuremath{t\bar{t}\rightarrow
    W^{+}bW^{-}\bar{b}\rightarrow \ell^{+}\nu_\ell b \ell^{-}\bar{\nu}_\ell \bar{b}}\xspace}

\newcommand{\xtrue}    {\ensuremath{y}\xspace}
\newcommand{\xreco}    {\ensuremath{x}\xspace}
\newcommand{\xeps}     {\ensuremath{q}\xspace}

\newcommand{\cme}      {\ensuremath{\sqrt{s}=1.96~\TeV}\xspace}
\newcommand{\GeV}{\ensuremath{\mathrm{Ge\kern-0.1em V}}\xspace}
\newcommand{\TeV}{\ensuremath{\mathrm{Te\kern-0.1em V}}\xspace}
\newcommand{\keV}{\ensuremath{\mathrm{ke\kern-0.1em V}}\xspace}

\newcommand{\met}       {\mbox{$\not\!p_T$}}

\newcommand{\pythia}    {\mbox{\textsc{pythia}}\xspace}
\newcommand{\alpgen}    {\mbox{\textsc{alpgen}}\xspace}
\newcommand{\mcatnlo}   {\mbox{\textsc{mc@nlo}}\xspace}
\newcommand{\vecbos}    {\mbox{\textsc{vecbos}}\xspace}
\newcommand{\geant}     {\mbox{\textsc{geant3}}\xspace}
\newcommand{\herwig}    {\mbox{\textsc{herwig}}\xspace}
\newcommand{\lumi}      {5.4~fb$^{-1}$}

\newcommand{\Eref}[1]{Eq.~(\ref{#1})}
\newcommand{\Figref}[1]{Fig.~\ref{#1}}
\newcommand{\Fref}[1]{Fi\-gure~\ref{#1}}

\newcommand{\Tref}[1]{Table~\ref{#1}}

\newcommand{\resultlliiab}{\ensuremath{\mt= 174.0 \pm 1.8 {\rm (stat)} \pm 2.4 {\rm (syst)}~\GeV }\xspace}
\newcommand{\worldaverage}{\ensuremath{\mt= 173.3 \pm 1.1~\GeV }\xspace}
\def\Journal#1#2#3#4{{#1} {\bf #2}, #3 (#4)}

\def\PLB{{Phys. Lett.}  B}
\def\PRL{Phys. Rev. Lett.}
\def\PRD{{Phys. Rev.} D}

\begin{document}
%\modulolinenumbers[1]
%\linenumbers

% the following line is for submission, including submission to the arXiv!!
\hspace{5.2in} \mbox{Fermilab-Pub-11/203-E}

\title{Precise measurement of the top quark mass in the dilepont channel at \dzero}
\affiliation{Universidad de Buenos Aires, Buenos Aires, Argentina}
\affiliation{LAFEX, Centro Brasileiro de Pesquisas F{\'\i}sicas, Rio de Janeiro, Brazil}
\affiliation{Universidade do Estado do Rio de Janeiro, Rio de Janeiro, Brazil}
\affiliation{Universidade Federal do ABC, Santo Andr\'e, Brazil}
\affiliation{Instituto de F\'{\i}sica Te\'orica, Universidade Estadual Paulista, S\~ao Paulo, Brazil}
\affiliation{Simon Fraser University, Vancouver, British Columbia, and York University, Toronto, Ontario, Canada}
\affiliation{University of Science and Technology of China, Hefei, People's Republic of China}
\affiliation{Universidad de los Andes, Bogot\'{a}, Colombia}
\affiliation{Charles University, Faculty of Mathematics and Physics, Center for Particle Physics, Prague, Czech Republic}
\affiliation{Czech Technical University in Prague, Prague, Czech Republic}
\affiliation{Center for Particle Physics, Institute of Physics, Academy of Sciences of the Czech Republic, Prague, Czech Republic}
\affiliation{Universidad San Francisco de Quito, Quito, Ecuador}
\affiliation{LPC, Universit\'e Blaise Pascal, CNRS/IN2P3, Clermont, France}
\affiliation{LPSC, Universit\'e Joseph Fourier Grenoble 1, CNRS/IN2P3, Institut National Polytechnique de Grenoble, Grenoble, France}
\affiliation{CPPM, Aix-Marseille Universit\'e, CNRS/IN2P3, Marseille, France}
\affiliation{LAL, Universit\'e Paris-Sud, CNRS/IN2P3, Orsay, France}
\affiliation{LPNHE, Universit\'es Paris VI and VII, CNRS/IN2P3, Paris, France}
\affiliation{CEA, Irfu, SPP, Saclay, France}
\affiliation{IPHC, Universit\'e de Strasbourg, CNRS/IN2P3, Strasbourg, France}
\affiliation{IPNL, Universit\'e Lyon 1, CNRS/IN2P3, Villeurbanne, France and Universit\'e de Lyon, Lyon, France}
\affiliation{III. Physikalisches Institut A, RWTH Aachen University, Aachen, Germany}
\affiliation{Physikalisches Institut, Universit{\"a}t Freiburg, Freiburg, Germany}
\affiliation{II. Physikalisches Institut, Georg-August-Universit{\"a}t G\"ottingen, G\"ottingen, Germany}
\affiliation{Institut f{\"u}r Physik, Universit{\"a}t Mainz, Mainz, Germany}
\affiliation{Ludwig-Maximilians-Universit{\"a}t M{\"u}nchen, M{\"u}nchen, Germany}
\affiliation{Fachbereich Physik, Bergische Universit{\"a}t Wuppertal, Wuppertal, Germany}
\affiliation{Panjab University, Chandigarh, India}
\affiliation{Delhi University, Delhi, India}
\affiliation{Tata Institute of Fundamental Research, Mumbai, India}
\affiliation{University College Dublin, Dublin, Ireland}
\affiliation{Korea Detector Laboratory, Korea University, Seoul, Korea}
\affiliation{CINVESTAV, Mexico City, Mexico}
\affiliation{FOM-Institute NIKHEF and University of Amsterdam/NIKHEF, Amsterdam, The Netherlands}
\affiliation{Radboud University Nijmegen/NIKHEF, Nijmegen, The Netherlands}
\affiliation{Joint Institute for Nuclear Research, Dubna, Russia}
\affiliation{Institute for Theoretical and Experimental Physics, Moscow, Russia}
\affiliation{Moscow State University, Moscow, Russia}
\affiliation{Institute for High Energy Physics, Protvino, Russia}
\affiliation{Petersburg Nuclear Physics Institute, St. Petersburg, Russia}
\affiliation{Instituci\'{o} Catalana de Recerca i Estudis Avan\c{c}ats (ICREA) and Institut de F\'{i}sica d'Altes Energies (IFAE), Barcelona, Spain}
\affiliation{Stockholm University, Stockholm and Uppsala University, Uppsala, Sweden}
\affiliation{Lancaster University, Lancaster LA1 4YB, United Kingdom}
\affiliation{Imperial College London, London SW7 2AZ, United Kingdom}
\affiliation{The University of Manchester, Manchester M13 9PL, United Kingdom}
\affiliation{University of Arizona, Tucson, Arizona 85721, USA}
\affiliation{University of California Riverside, Riverside, California 92521, USA}
\affiliation{Florida State University, Tallahassee, Florida 32306, USA}
\affiliation{Fermi National Accelerator Laboratory, Batavia, Illinois 60510, USA}
\affiliation{University of Illinois at Chicago, Chicago, Illinois 60607, USA}
\affiliation{Northern Illinois University, DeKalb, Illinois 60115, USA}
\affiliation{Northwestern University, Evanston, Illinois 60208, USA}
\affiliation{Indiana University, Bloomington, Indiana 47405, USA}
\affiliation{Purdue University Calumet, Hammond, Indiana 46323, USA}
\affiliation{University of Notre Dame, Notre Dame, Indiana 46556, USA}
\affiliation{Iowa State University, Ames, Iowa 50011, USA}
\affiliation{University of Kansas, Lawrence, Kansas 66045, USA}
\affiliation{Kansas State University, Manhattan, Kansas 66506, USA}
\affiliation{Louisiana Tech University, Ruston, Louisiana 71272, USA}
\affiliation{Boston University, Boston, Massachusetts 02215, USA}
\affiliation{Northeastern University, Boston, Massachusetts 02115, USA}
\affiliation{University of Michigan, Ann Arbor, Michigan 48109, USA}
\affiliation{Michigan State University, East Lansing, Michigan 48824, USA}
\affiliation{University of Mississippi, University, Mississippi 38677, USA}
\affiliation{University of Nebraska, Lincoln, Nebraska 68588, USA}
\affiliation{Rutgers University, Piscataway, New Jersey 08855, USA}
\affiliation{Princeton University, Princeton, New Jersey 08544, USA}
\affiliation{State University of New York, Buffalo, New York 14260, USA}
\affiliation{Columbia University, New York, New York 10027, USA}
\affiliation{University of Rochester, Rochester, New York 14627, USA}
\affiliation{State University of New York, Stony Brook, New York 11794, USA}
\affiliation{Brookhaven National Laboratory, Upton, New York 11973, USA}
\affiliation{Langston University, Langston, Oklahoma 73050, USA}
\affiliation{University of Oklahoma, Norman, Oklahoma 73019, USA}
\affiliation{Oklahoma State University, Stillwater, Oklahoma 74078, USA}
\affiliation{Brown University, Providence, Rhode Island 02912, USA}
\affiliation{University of Texas, Arlington, Texas 76019, USA}
\affiliation{Southern Methodist University, Dallas, Texas 75275, USA}
\affiliation{Rice University, Houston, Texas 77005, USA}
\affiliation{University of Virginia, Charlottesville, Virginia 22901, USA}
\affiliation{University of Washington, Seattle, Washington 98195, USA}
\author{V.M.~Abazov} \affiliation{Joint Institute for Nuclear Research, Dubna, Russia}
\author{B.~Abbott} \affiliation{University of Oklahoma, Norman, Oklahoma 73019, USA}
\author{B.S.~Acharya} \affiliation{Tata Institute of Fundamental Research, Mumbai, India}
\author{M.~Adams} \affiliation{University of Illinois at Chicago, Chicago, Illinois 60607, USA}
\author{T.~Adams} \affiliation{Florida State University, Tallahassee, Florida 32306, USA}
\author{G.D.~Alexeev} \affiliation{Joint Institute for Nuclear Research, Dubna, Russia}
\author{G.~Alkhazov} \affiliation{Petersburg Nuclear Physics Institute, St. Petersburg, Russia}
\author{A.~Alton$^{a}$} \affiliation{University of Michigan, Ann Arbor, Michigan 48109, USA}
\author{G.~Alverson} \affiliation{Northeastern University, Boston, Massachusetts 02115, USA}
\author{G.A.~Alves} \affiliation{LAFEX, Centro Brasileiro de Pesquisas F{\'\i}sicas, Rio de Janeiro, Brazil}
% ...
%
% visitor_addresses.tex                        6 April 2011
%  available symbols are:
%  $\ast, \dag, \ddag, \S, \P, $\|$, $\ast\ast$, \dag\dag, \ddag\ddag ,\#
%
\collaboration{The D0 Collaboration\footnote{with visitors from
%{alton}
$^{a}$Augustana College, Sioux Falls, SD, USA,
%{burdin}
$^{b}$The University of Liverpool, Liverpool, UK,
%{haas,partridge}
$^{c}$SLAC, Menlo Park, CA, USA,
%{hesketh}
$^{d}$University College London, London, UK,
%{luna-garcia}
$^{e}$Centro de Investigacion en Computacion - IPN, Mexico City, Mexico,
%{podesta-lerma}
$^{f}$ECFM, Universidad Autonoma de Sinaloa, Culiac\'an, Mexico,
and 
%{weber}
$^{g}$Universit{\"a}t Bern, Bern, Switzerland.
%{garcia-guerra}
%$^{?}$UPIITA-IPN, Mexico City, Mexico,
%{hooper}
%$^{?}$Visitor from Bradley University, Peoria, IL, USA.
%{kozminski}
%$^{?}$}Visitor from Lewis University, Romeoville, IL, USA.
%{deceased}
%$^{\ddag}$Deceased.
}} \noaffiliation
\vskip 0.25cm
       % D0 authors (remove the first 3 lines
                             % of this file prior to submission, they
                             % contain a time stamp for the authorlist)
                             % (includes institutions and visitors)
\date{April 28, 2011}

\begin{abstract}
We measure the top quark
mass (\mt) in \ppbar collisions at a center of mass energy \cme 
using dilepton \fulldecay events, where $\ell$ denotes an 
electron, a muon, or a tau that decays leptonically. The data 
correspond to an integrated luminosity of \lumi\ collected
with the \dzero detector at the Fermilab Tevatron Collider. 
We obtain \resultlliiab, which is in
agreement with the current world average \worldaverage. 
This is currently the most precise measurement of \mt
in the dilepton channel.

\end{abstract}

\pacs{14.65.Ha}
\maketitle

%\section{\label{sec:level1}First-level heading}
% sections are not used for PRL papers

The measurement of the properties of the top quark has been a 
major goal of the Fermilab Tevatron Collider experiments since 
its discovery in 1995~\cite{cdf-top-discovery, d0-top-discovery}.
As the heaviest known elementary particle, the top quark may play a 
special role in the mechanism of electroweak symmetry breaking.
A precise measurement of its mass (\mt) is of particular importance, since, 
combined with the measurement of the $W$ boson mass, it provides
an indirect constraint on the mass of the Higgs boson in the standard 
model (SM), and can also constrain possible extensions of the SM.

We present a new measurement of the top quark mass in the dilepton
channel (\ee, \emu, \mumu) in \fulldecay\ events, where $\ell$ 
denotes an electron, a muon or a tau decaying leptonically, 
using the matrix element method. 
The first measurement of \mt based on this method was performed
in the lepton+jets channel by the \dzero experiment~\cite{top-ME-ljets-D0}. 
The CDF Collaboration has applied the matrix element approach to determine \mt in the 
dilepton and all-hadronic final states~\cite{top-ME-ll-CDF,top-ME-all-CDF}, 
obtaining a mass precision of $4.0~\GeV$ for dilepton 
events~\cite{top-ME-ll-CDF}.
The measurement of \mt in the dilepton channel has also been 
carried out by using other techniques~\cite{cdf-top-mass-ll-other,
cdf-top-mass-ll-other2,cdf-top-mass-ll-other3,cdf-top-mass-ll-other4,
d0-top-mass-ll-other,d0-top-mass-ll-other2}, reaching a precision of $3.7~\GeV$.
We report a measurement based on data
collected by the \dzero detector, corresponding to \lumi\ of 
integrated luminosity from \ppbar\ collisions at \cme. 

The \dzero detector has a central tracking system, consisting of a
silicon microstrip tracker and a central fiber tracker,
both located within a 1.9~T superconducting solenoidal
magnet~\cite{run2det}, with the design providing tracking and
vertexing at pseudorapidities $|\eta|<3$~\cite{eta}. 
The liquid-argon and uranium calorimeter has a
central section covering pseudorapidities $|\eta|$ up to
$\approx 1.1$ and two end calorimeters that extend coverage
to $|\eta|\approx 4.2$, with all three housed in separate
cryostats~\cite{run1det}. 
A muon system outside the calorimeters covers $|\eta|<2$ and consists
of a layer of tracking detectors and scintillation trigger
counters in front of 1.8~T toroids, followed by two similar layers
after the toroids~\cite{run2muon}.

Despite the small branching fraction of this final state and the presence
of two neutrinos in each event, the measurement of \mt in the dilepton channel
is interesting because the lower background and the smaller jet multiplicity
relative to the lepton+jets channel result in a reduced sensitivity
to the ambiguity from combining jets in the reconstruction of \mt. 
The dilepton measurement therefore complements the results from other final 
states. Moreover, significant differences in measured values of
\mt in different \ttbar decay channels can be indicative of the presence
of physics beyond the SM~\cite{Kane:1996ny}.

As the SM predicts top quarks to decay almost 100\% of the
time into a $W$ boson and a $b$ quark, \ttbar events are classified
according to the decays of the $W$ boson. In the dilepton channel, 
both $W$ bosons decay leptonically, \wlnu~\cite{chargeconjugation}
with $\ell=e,\mu$ or $\tau$. We analyze the events 
characterized by two leptons \ee, \emu, or \mumu,
with a large transverse momenta (\pt), large imbalance in transverse 
momentum from the undetected neutrinos ($\met$), and two high-\pt jets 
from the $b$ quarks. The \wtaunu decays contribute 
through secondary \taulnunu transitions. For the $ee$ and $\mu\mu$ analysis, 
we consider events selected by a set of 
single-lepton triggers. For the $e\mu$ channel, we use a mixture of single 
and multilepton triggers and lepton+jet triggers. Dilepton \ttbar\ events are 
required to have at least two oppositely charged, isolated leptons with 
$\pt > 15~\GeV$, and either $| \eta |  < 1.1 $ or  $ 1.5 < |\eta| < 2.5 $ 
for electrons and $|\eta| < 2$ for muons.
If more than one lepton-pair combination is found in an event, only the pair 
with the largest sum in scalar \pt is used. 
Events must have at least two jets with $\pt > 20~\GeV$ and $|\eta | < 2.5$,
well separated from the selected electrons. 
No explicit $b$-jet identification is required in this analysis.
The main sources of background in the dilepton channel are
Drell-Yan and $Z$ boson production (\zll),
diboson production ($WW, WZ, ZZ$), and instrumental background
that originates from limited detector resolution and lepton misidentification.
In the \ee channel, the discrimination between the \ttbar 
signal and background improves by requiring 
a large significance of the measured $\met$, %($\metsig$),
which is defined through a likelihood discriminant constructed from the ratio 
of $\met$ to its uncertainty~\cite{metsig}.
In the \mumu channel, we require, in addition, $\met > 40~\GeV$.
In the \emu channel, the requirement $H_T > 115~\GeV$,
where $H_T$ is the scalar sum of the transverse momenta of the leading 
lepton and the two leading jets, rejects most of the contribution
from \taulnunu. 
The above selections minimize the expected statistical uncertainty on \mt. 
In total, we select 479 candidate events with 73, 266, and 140 events, 
respectively, in the \ee, \emu, and  \mumu channels, of which about 
$13\pm 5$, $48\pm 15$, and $56\pm 15 $
events, respectively, are expected to arise from the background. 

The matrix element method is based on the probability for a given event to 
resemble a signal, which depends on the value of \mt, or a background, 
which is usually independent of \mt.
Assuming that the different physics processes leading to the same final 
state do not interfere, the event probability can be written as 
the sum of probabilities from all possible contributions.
In practice, because the matrix element method requires significant computing time, only 
the dominant background is taken into account, and the total event probability
is given by 
\begin{equation}
\pevt = \fttbar \pttbar(\xreco;\mt)+(1-\fttbar)\pzjj(\xreco),
\label{eq:pevt}
\end{equation}

where \fttbar is the fraction of \ttbar\ events, \pttbar and
\pzjj are the signal and background probability densities, respectively, 
\mt is the assumed top quark mass, and \xreco reflects the
observed kinematic variables, i.e., the four-momenta of the measured jets and leptons.
In the \ee, \mumu, and \emu channels, 
\zjj events with \zee, \zmumu and \ztautauemu 
are the dominant source of background. The second leading background,
from misidentified leptons, is approximately a factor of 3 smaller.
While neglecting the other background probabilities leads to some bias, the calibration
procedure described below allows us to correct for these and other limitations of the model.
%There is no bias expected from neglecting other 
%background probabilities, as the analysis is calibrated using all significant 
%sources of background, which provides a way to correct for the limitations 
%of the model, as described below.   
 
The leading-order (LO) matrix element for $\qqbar\rightarrow\fulldecay$ is 
used to compute the \ttbar probability density.    
For each final state \xtrue of the six produced partons, 
%$i$, with four-momenta $p_i, i=1,\dots,6$, 
the signal probability is given by 
\begin{equation}
\begin{split}
\pttbar(\xreco;\mt) & = 
 \frac{1}{\sigmaobs(\mt)} \cdot \\
& \sum_{i=1}^{8} \int  %_{-\infty}^{\infty}
~{\rm d}\xeps_1~{\rm d}\xeps_2~f_{\rm PDF}(\xeps_1) ~f_{\rm PDF}(\xeps_2) \cdot \\
& \frac{(2 \pi)^{4} \left|M(\xtrue;\mt)\right|^{2}}
       {\xeps_1 \xeps_2 s}
 ~     {\rm d}\Phi_{6}
 ~W(\xreco,\xtrue)~W(\ptttbar) \ ,
\end{split} 
\label{eq:psgn}
\end{equation}

where $\xeps_1$ and $\xeps_2$ denote the momentum fractions of the
incident quarks in the proton and antiproton, respectively, $f_{\rm PDF}$ are the
parton distribution functions (PDF) for finding a parton of a given flavor
and longitudinal momentum fraction in the proton or antiproton 
(in this analysis we use the CTEQ6L1 PDF~\cite{bib:CTEQ6L1PDF}), 
$s$ is the square of the energy in the $q\bar{q}$ rest frame, 
$M(\xtrue)$ is the leading-order matrix element~\cite{bib:MAHLONPARKE} and
${\rm d}\Phi_{6}$ is an element of the 6-body phase space.   
Detector resolution is taken into account through a transfer
function $W(\xreco,\xtrue)$ that describes the probability of the partonic final
state $\xtrue$ to be measured as $\xreco$. 
The finite transverse momentum of the \ttbar\ system is
accounted for through an integration over its probability distribution, 
which is derived from parton-level simulated events using 
\alpgen~\cite{alpgen}, employing \pythia~\cite{pythia}
for parton showers and hadronization.
%The differential cross section is normalized to the observable cross 
%section \sigmaobs which corresponds to the cross section after selection cuts.
As the angular resolution of the jets and leptons, as well as the electron 
energy resolution, are sufficiently well determined, there is no need to 
introduce resolution functions.
By taking into account energy and momentum conservation,
\Eref{eq:psgn} can be reduced to an
integration over the energies associated with the $b$ quarks, the
lepton-neutrino invariant masses squared, the differences between neutrino transverse 
momenta, the transverse momentum of the \ttbar
system, and the radii of curvature ($\pt^{-1}$) of muons. 
The sum %$\sum_{i=1}^{8}$ 
runs over both possible
jet-parton assignments and over up to two real solutions
for each neutrino energy~\cite{bib:MENIMPAPER}. 
The normalization factor \sigmaobs is 
the product of the LO cross section 
and the mean efficiency of the final selections. 
A transfer function $W(\xreco,\xtrue)$ is used
for each jet and each muon in the final state. 
The jet energy resolution is parametrized as the sum of 
two Gaussian functions, with parameters depending linearly 
on parton energies, while the resolution in muon
$\pt^{-1}$ is described by a single Gaussian function.     
All parameters in $W(\xreco,\xtrue)$ are determined from Monte Carlo (MC) 
\ttbar\ events, tuned to match the resolutions observed in the data. 
%The electron resolution is sufficiently good to have no effect on the expected
%uncertainty on \mt if it is ignored in the transfer function.

To take account of all background processes and to provide
a correct statistical sampling of possible spin, flavor,
and color configurations, the background probability \pzjj is
calculated by using \vecbos~\cite{bib:vecbos}.  
Since \ztautau decay is not modeled in \vecbos, an additional transfer
function in the \emu channel is used
to describe the energy of the final state lepton relative 
to the initial $\tau$ lepton, derived from parton-level
information~\cite{bib:MENIMPAPER}. The direction of the final state
lepton is assumed to be close to that of the $\tau$ lepton,
since only in such cases is the lepton from the $\tau$ decay sufficiently 
energetic to pass the \pt selection. 
For the \ztautauemujj probability, the energy fractions for final state
leptons are sampled according to this $\tau$ transfer function.
The jet and charged-lepton 
directions are assumed to be well-measured, and each kinematic
solution is weighted according to the \pt\ of the \zjj system.
The integration of the probability for \zjj is
performed over the energies of the two partons that lead to the jets.  
Both possible assignments of jets to quarks are considered.

To calculate the signal and background probability densities, a MC-based
integration of \Eref{eq:psgn} is performed and \mt 
is changed in steps of 2.5~\GeV over a range of 30~\GeV.
For each mass hypothesis, a likelihood function
$\ptot(\mt,\fttbar)$ is defined by the product of individual
event probabilities \pevt, and the signal fraction \fttbar is determined
by minimizing $-\ln\ptot$. %at that mass point. 
Finally, the most likely value of \mt and its uncertainty are extracted from 
a fit of $\ptot(\mt)$ to a Gaussian form near its maximum by using the 
value of \fttbar found in the previous step.

\begin{figure}
\setlength{\unitlength}{1.1cm}
\begin{picture}(8.0,4.0)
\put(0.0,0.2){\includegraphics[scale=0.245]{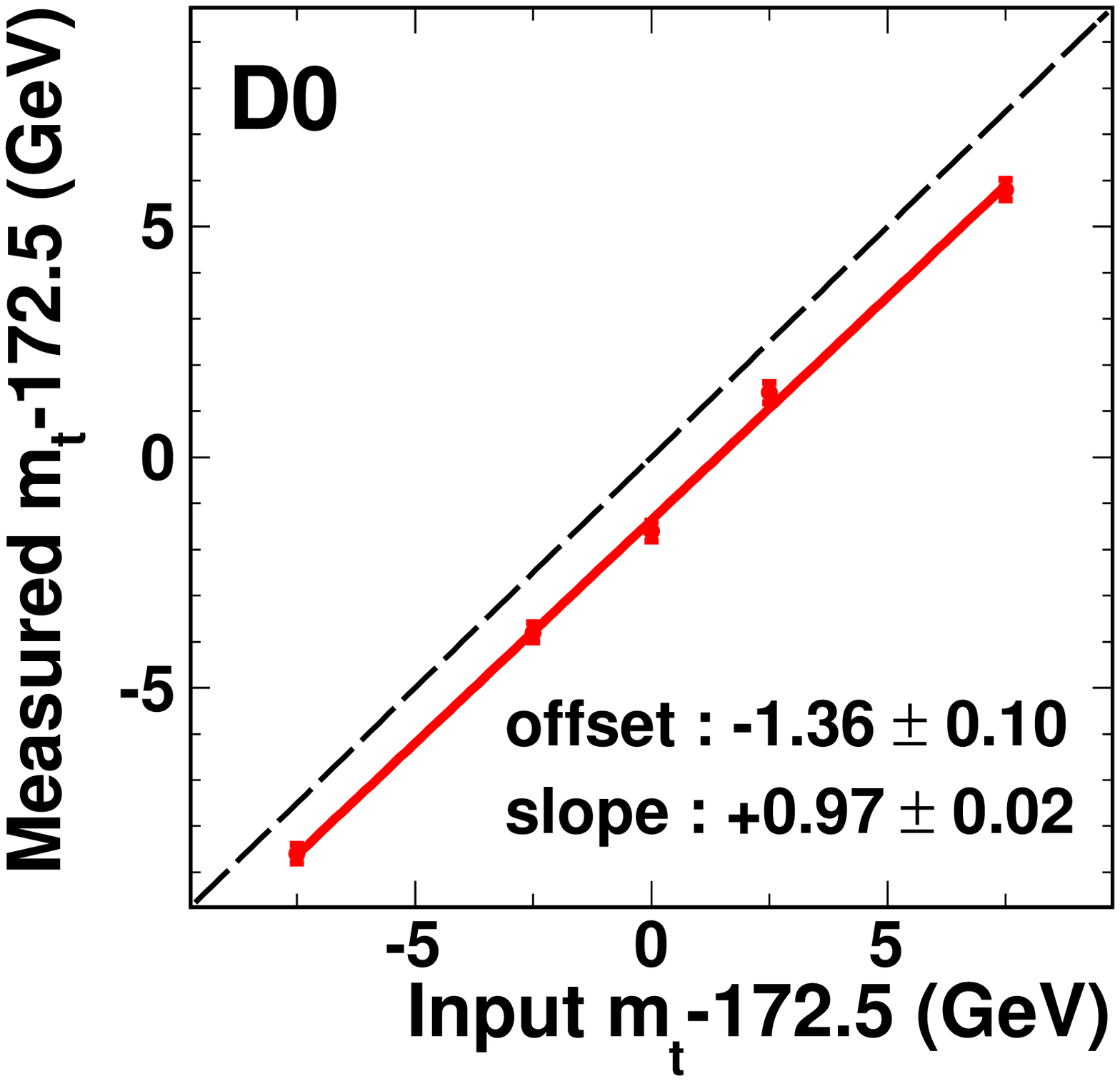}}
\put(4.0,0.2){\includegraphics[scale=0.245]{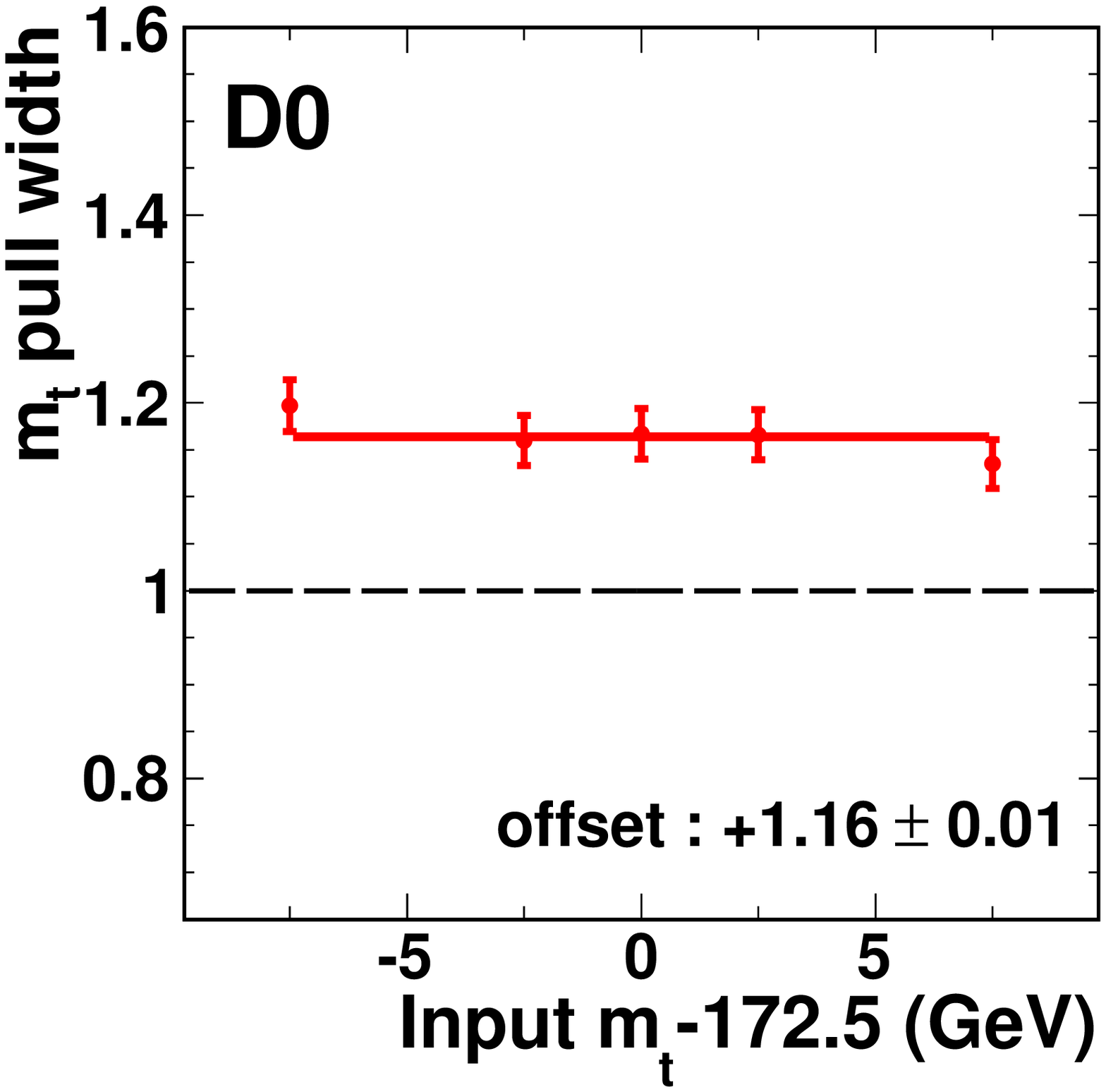}}
\put(0.9,3.1){\bf(a)}
\put(4.9,3.1){\bf(b)}
\end{picture}
\caption{\label{fig:calibration} (a) Mean values of \mt and (b)
pull width from sets of 1000 pseudoexperiments as a function of 
input \mt for the combined dilepton channels.
The dashed lines represent the ideal response in (a), where
the extracted mass is identical to the input mass, and in (b), 
where the statistical uncertainty requires no correction.}
\end{figure}

\begin{figure}
\setlength{\unitlength}{1.1cm}
\begin{picture}(8.0,4.0)
\put(0.0,0.2){\includegraphics[scale=0.245]{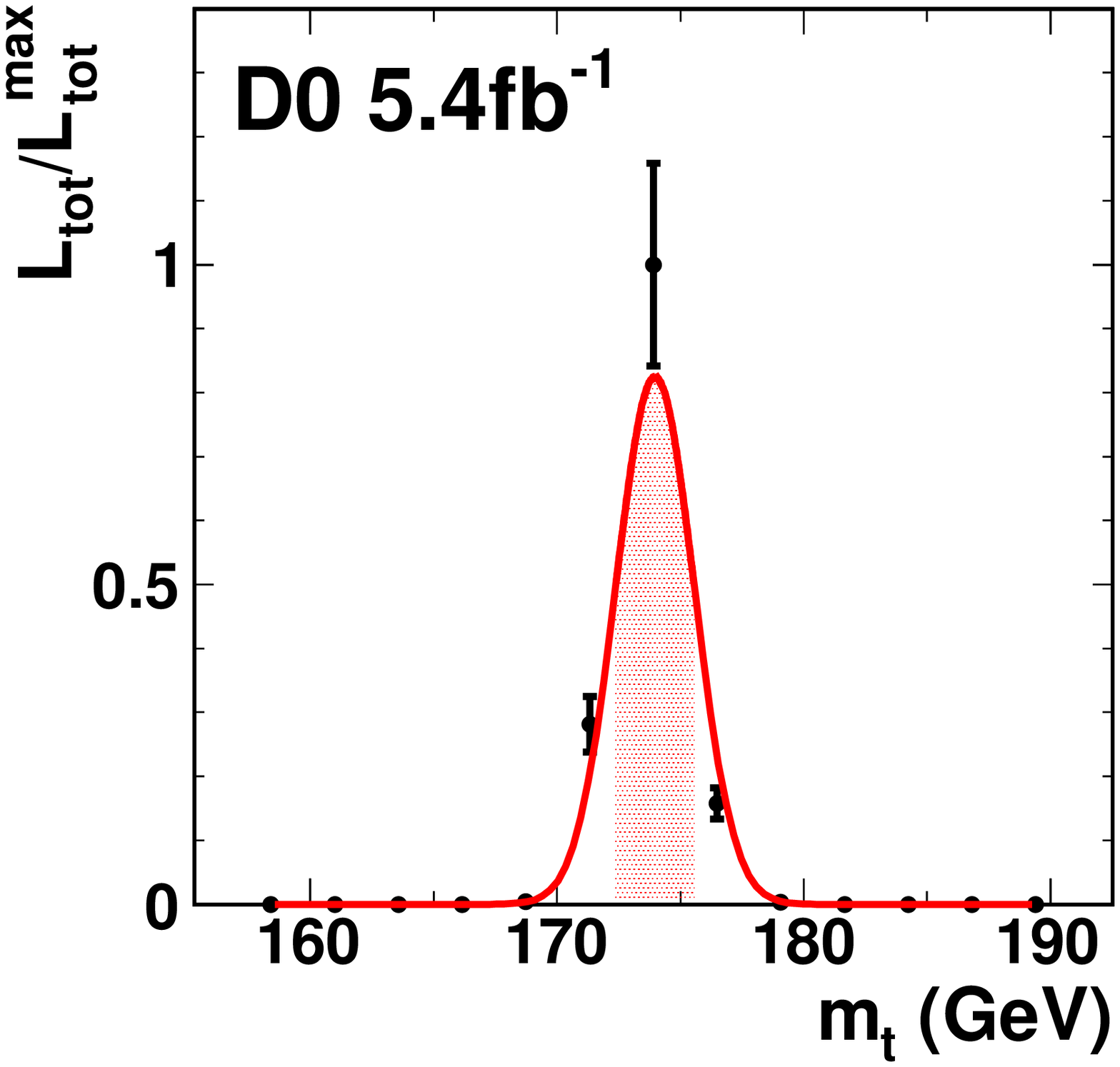}}
\put(4.0,0.2){\includegraphics[scale=0.245]{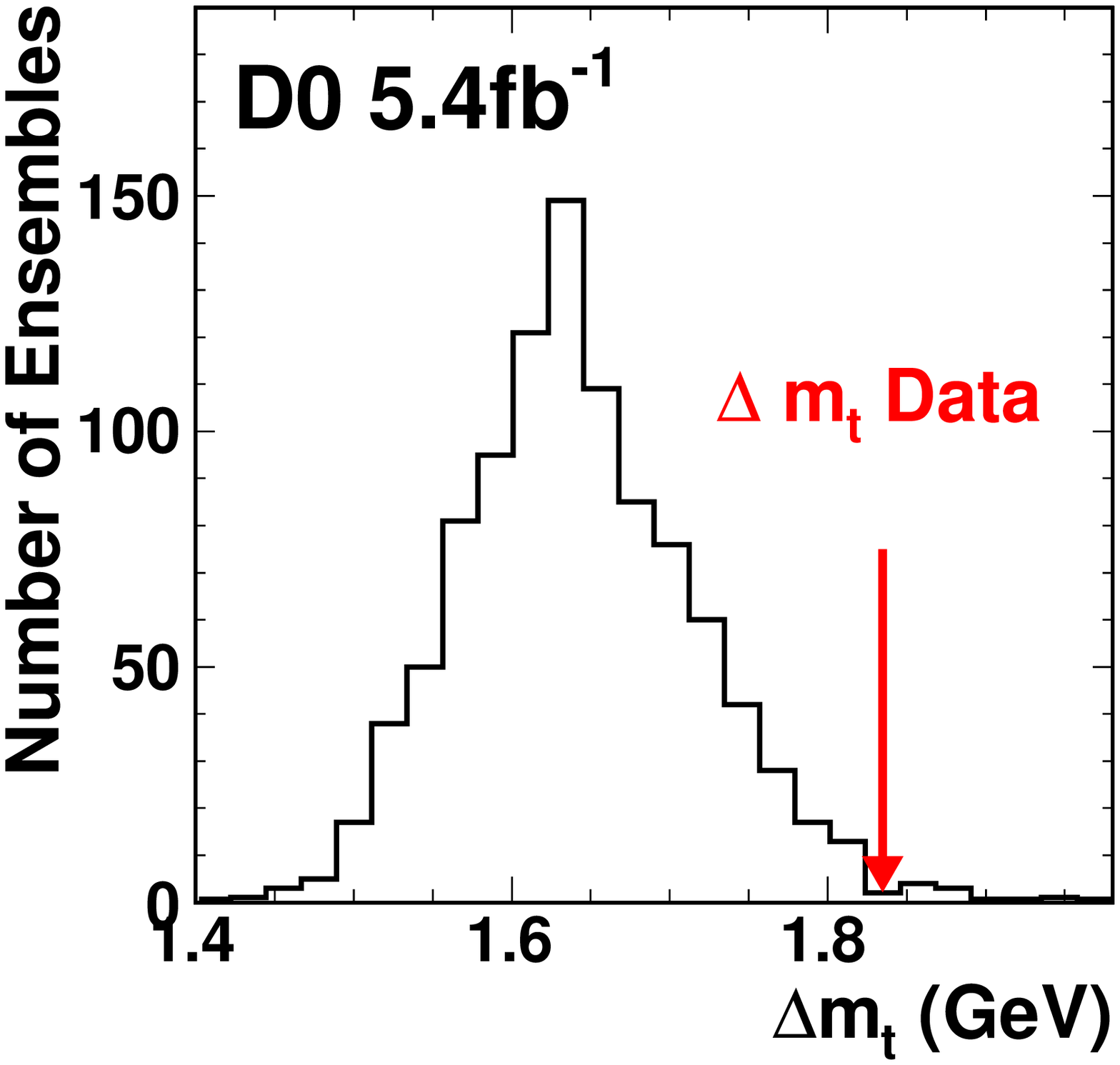}}
\put(0.9,3.1){\bf(a)}
\put(4.9,3.1){\bf(b)}
\end{picture}
\caption{\label{fig:resultdata} Combined for all channels: (a) Calibrated 
and normalized likelihood for the data as a function of \mt with the best estimate 
as well as 68\% confidence level region marked by the shaded area 
and in (b) the expected distribution of uncertainties with the measured 
uncertainty indicated by the arrow. As the top quark mass is measured to 
be $\mt=174.0~\GeV$, the expected distribution in (b) is shown for the 
closest input mass $\mt=175~\GeV$ used in the pseudoexperiment.}
\end{figure}

To check for any bias caused by approximations of the method, such
as the use of the LO matrix element for \pttbar or from neglecting
backgrounds other than \zjj, the measurement
is calibrated by using MC events generated with \alpgen+~\pythia.
All events are processed through a full
\geant \cite{bib:geant} detector simulation, followed by the same
reconstruction and analysis chain as used for the data.
Effects from additional \ppbar\ interactions are simulated by overlaying 
the data from random \ppbar\ crossings over the MC events.
Five \ttbar MC samples are generated with input top quark masses of $\mt=165$,
170, 172.5, 175, and 180~\GeV. Probabilities for the \ttbar\ signal and 
for \zll, diboson and instrumental backgrounds, are 
used to form randomly drawn pseudoexperiments.
The total number of events in each pseudoexperiment is fixed 
to the number of events in the data for the combined dilepton channels. 
The signal and background fractions are fluctuated according to multinomial 
statistics around the fractions determined from the measured \ttbar\ cross 
section in the separate channels~\cite{bib:d0_spin}.
The mean values of \mt measured in 1000 pseudoexperiments
as a function of the input \mt are shown in
\Figref{fig:calibration}(a). The deviation from the ideal response, where
the extracted mass is equal to the input mass, is caused both by  
the presence of backgrounds without a corresponding matrix element in the event 
probability and by approximations in the calculation of the \zjj probabilities.  
For the case of background-free pseudoexperiments, no difference is observed.  
%The pull width, defined as the deviation of \mt in a single pseudoexperiment 
%from the mean of all 1000 measurements 
The width of the distribution of the pulls ("pull width"), defined as
the mean deviation of \mt in single pseudoexperiments from the mean for
all 1000 values at a given input \mt, 
in units of the measured uncertainty per pseudoexperiment, is shown in
\Figref{fig:calibration}(b). The statistical uncertainty measured in the data 
is corrected for the deviation of the pull width from unity.
The calibrated value of \mt from the fit to the data 
is shown in \Figref{fig:resultdata}(a). \Fref{fig:resultdata}(b) compares the 
measured uncertainty for \mt with the 
distribution of expected uncertainties in pseudoexperiments at $\mt=175~\GeV$.
The difference between the observed and median expected uncertainty is not 
statistically significant. We also note that, 
%it can be explained by effects 
%that are covered by systematic uncertainties.
%For example
when we change the signal to background ratio within
uncertainties, the expected uncertainty generally increases and agrees well with the 
observation.

Systematic uncertainties on the measurement of \mt 
can be divided into three categories. The first involves uncertainties from
modeling of the detector, such as the uncertainty on the energy scale of 
light-quark jets and the uncertainty in the relative calorimeter response to $b$
and light-quark jets, as well as in the energy resolution for jets, muons,
and electrons. The second category is related to the
modeling of \ttbar\ production. This includes possible differences in the 
amount of initial and final state radiation, effects from next-to-leading-order 
contributions and different hadronization models, color reconnection, 
and modeling of $b$-quark fragmentation as well as uncertainties from 
the choice of PDF.
The third category comprises effects from
calibration, such as the uncertainties in the calibration function shown in 
\Figref{fig:calibration}(a), and from variations in signal and background 
contributions in the pseudoexperiments. Contributions to the total systematic
uncertainty in the measurement of \mt are summarized in \Tref{tab:systematics}.

The dominant systematic uncertainty 
%in the dilepton channel 
arises from the 
different detector response of light and $b$-quark jets.
It accounts for the different calorimeter response of single
pions in the data and MC simulation and the different fractions of single pions in light and 
$b$-quark jets. The relative uncertainty of the response has been evaluated 
to be 1.8\% leading to a shift of 1.6~\GeV in \mt.
%This difference is propagated to the energy scale (JES)
%of jets arising from b quarks.
%The dominant systematic uncertainty in the dilepton channel arises from 
%the difference between the nominal inclusive response derived for light quark jets  
%and the response of jets from $b$ quarks. 
%It accounts for the differences in the calorimeter response to electromagnetic
%and hadronic showers and the uncertainty on the electromagnetic and hadronic 
%energy faction in jets from b quarks. 
The next important uncertainty comes from uncertainties 
in the jet energy scale (JES) of light quarks.
This JES is calibrated by using $\gamma+$jets
events~\cite{jes}. More than 80\% of the JES uncertainty is due to the understanding
of the detector response and the showering of jets.  
The total uncertainty typically adds up to about 1.5\% per jet, 
which translates into an uncertainty on \mt of 1.5~\GeV. 
The main uncertainty from modeling \ttbar\ production is
from higher-order effects and hadronization. It 
is evaluated by using \ttbar events generated
with \mcatnlo~\cite{bib:mcatnlo} and evolved in \herwig~\cite{Corcella:2000bw}.
The next leading uncertainty on modeling \ttbar\ arises from the
description of $b$-quark fragmentation. It is derived by comparing
the extracted \mt for the default measurement with the result using a
reweighting of the default MC samples to a Bowler scheme tuned to LEP 
or SLD data ~\cite{bib:bfraglepsld}.
The largest difference is quoted as the uncertainty. 

In summary, we have presented a measurement of the top quark mass in 
the \fulldecay channel using the matrix element method. Based on an
integrated luminosity of \lumi\ collected by the \dzero Collaboration, 
the top quark mass is found to be
\begin{equation}
\resultlliiab . 
\label{eq:result}
\end{equation}

This measurement is in good agreement with the current world average 
\worldaverage~\cite{bib:waverage}. Its total uncertainty of $3.1~\GeV$ 
corresponds to a 1.8~\% accuracy and represents the most precise measurement 
of \mt from dilepton \ttbar\ final states.

\begin{table}
\begin{center}
\caption{Summary of systematic uncertainties on the measurement of 
\mt in dilepton events.}
\begin{tabular}{lc}
\hline
\hline
 Source & Uncertainty (\GeV) \\
\hline
\it{Detector modeling:}\\
$b$/light jet response & $\pm 1.6$ \\
JES                  & $\pm 1.5$ \\
Jet resolution       & $\pm 0.3$ \\
Muon resolution      & $\pm 0.2$  \\
Electron \pt scale   & $\pm 0.4$ \\
Muon \pt scale       & $\pm 0.2$ \\
ISR/FSR              & $\pm 0.2$ \\
\it{Signal modeling:}\\
Higher order and hadronization & $\pm 0.7$ \\
Color reconnection   & $\pm 0.1$  \\
$b$-quark modeling   & $\pm 0.4$ \\
PDF uncertainty      & $\pm 0.1$ \\
\it{Method:}\\
MC calibration       & $\pm 0.1$ \\
Signal fraction      & $\pm 0.5$ \\
\hline
Total		     & $\pm 2.4$ \\
\hline
\hline
\end{tabular}
\label{tab:systematics}
\end{center}
\end{table}

% acknowledgement.tex                             6 October 2010
%
We thank the staffs at Fermilab and collaborating institutions,
and acknowledge support from the
DOE and NSF (USA);
CEA and CNRS/IN2P3 (France);
FASI, Rosatom and RFBR (Russia);
CNPq, FAPERJ, FAPESP and FUNDUNESP (Brazil);
DAE and DST (India);
Colciencias (Colombia);
CONACyT (Mexico);
KRF and KOSEF (Korea);
CONICET and UBACyT (Argentina);
FOM (The Netherlands);
STFC and the Royal Society (United Kingdom);
MSMT and GACR (Czech Republic);
CRC Program and NSERC (Canada);
BMBF and DFG (Germany);
SFI (Ireland);
The Swedish Research Council (Sweden);
and
CAS and CNSF (China).
%
   % input acknowledgement


\begin{thebibliography}{99}
\bibitem{eta}
  The pseudorapidity $\eta$ is defined relative to the center of the detector 
as $\eta = -\ln [\tan(\theta/2)]$ where $\theta$ is the polar angle with 
respect to the proton beam direction.
\bibitem{chargeconjugation}
Throughout this Letter, charge conjugated processes are included implicitly.
\end{thebibliography}
\end{document}